\documentclass[prb,twocolumn,amssymb,showpacs]{revtex4}
\usepackage{graphicx}
\usepackage{amsmath}
\usepackage{txfonts}

\newlength{\figwidth}
 \divide\figwidth by 2
\newcommand{\tbti}{Tb$_{2}$Ti$_{2}$O$_7$~}

\newcommand{\mub}{$\mu_B$~}

\begin{document}

\title{Field induced  spin ice like orders in spin liquid Tb$_2$Ti$_2$O$_7$}
\author{H. Cao$^1$, A. Gukasov$^1$, I. Mirebeau$^1$, P. Bonville$^2$,
and G. Dhalenne$^3$.
\\
}
\address{
$^1$Laboratoire L\'eon Brillouin, CEA-CNRS, CE-Saclay, 91191
Gif-sur-Yvette, France.}
\address {$^2$Service de Physique de l'Etat Condens\'e,
CEA-CNRS, CE-Saclay,  91191 Gif-Sur-Yvette, France.}
\address {$^3$Laboratoire de Physico-Chimie de l'Etat Solide Universit\'e Paris-Sud,
91405 Orsay, France.}

\begin{abstract}
We have studied the field induced magnetic structures in the spin
liquid Tb$_2$Ti$_2$O$_7$, in a wide temperature
(0.3$<$$T$$<$270~K) and field (0$<$$H$$<$7~T) range, by single
crystal neutron diffraction with $\bf{H}$ // [110] axis. We
combined unpolarized neutron data with polarized ones, analyzed
within the local susceptibility model. A ferromagnetic-like
structure with $\bf{k}$ = 0 propagation vector is induced, whose
local order at low field and low temperature is akin to spin ice.
The four Tb ions separate in $\alpha$ and $\beta$ chains having
different values of the magnetic moments, which is quantitatively
explained by taking the crystal field anisotropy into account.
Above 2~T and below 2~K, an antiferromagnetic-like structure with
$\bf{k}$ = (0,0,1) is induced besides the $\bf{k}$ = 0 structure.
It shows a reentrant behavior and extends over a finite length
scale. It occurs together with a broadening of the nuclear peaks,
which suggests a field induced distortion and magnetostriction
effect.

\end{abstract}

 \maketitle

 Geometrical frustration now attracts considerable interest,
 as being a possible
 tool for tuning several physical properties concomitantly.
 The pyrochlore lattice of corner sharing tetrahedra offers the best model of this
 frustration in three dimensions, leading to short range magnetic
 orders such as spin liquids, spin ices and spin glasses without chemical disorder.
 R$_2$Ti$_2$O$_7$ pyrochlores where R=Dy or Ho are model spin ices\cite{Bramwell012}, with a ground state entropy akin
 to that of real ice\cite{Ramirez99}. Ho or Dy magnetic
moments interact via ferromagnetic first neighbor interactions and
are constrained to point along their local $<$111$>$ Ising axes.
The local spin ice structure, with 6-fold degeneracy, consists of
two spins pointing in and two out of each tetrahedron.

The spin ice degeneracy is lifted by a magnetic field $H$, leading
to original magnetic transitions, which depend on the orientation
of $\bf{H}$ with respect to the local $<$111$>$ trigonal
axes\cite{Harris98}. With $\bf{H}$ // [111],
magnetization plateau  
and liquid-gas transition
occur\cite{Matsuhira01,Sakakibara03,Tabata06}, providing the first
evidence of magnetic monopoles\cite{Castelnovo08}. Applying
$\bf{H}$ // [100] yields the first magnetic
example\cite{Jaubert08} of the Kasteleyn transition known in
polymers. With $\bf{H}$
// [110], the lattice divides into $\alpha$ and $\beta$ chains\cite{Hiroi03,Ruff05,Fennell05},
with different angles between $\bf{H}$ and the local Ising axes.

In Tb$_2$Ti$_2$O$_7$, the crystal field (CF) anisotropy is weaker
than in canonical spin ices, and ferromagnetic (F) and
antiferromagnetic (AF) first neighbor interactions nearly
compensate. This widely studied spin
liquid\cite{Gardner99,Enjalran04}, where short range correlated
moments fluctuate down to 50 mK, was
 also called a $"$quantum spin ice$"$\cite{Molavian07}. Its magnetic ground
state (GS) is very sensitive to perturbations. Lattice expansion
induced by substituting Sn for Ti yields a long range ``ordered
spin ice'' in Tb$_2$Sn$_2$O$_7$ \cite{Mirebeau05}. An AF long
range order (LRO) is induced under applied stress and/or magnetic
field\cite{Mirebeau02,Mirebeau04,Rule06}.

 Up to now, the complex field induced magnetic structures were not studied
 precisely in Tb$_2$Ti$_2$O$_7$. 
 We have combined high accuracy neutron
 diffraction experiments with both polarized and unpolarized neutrons,
 in a wide temperature (0.3$<$$T$$<$270~K) and  field (0$<$$H$$<$7 T) range,
to investigate them in detail.
 The polarized neutron technique, newly used for such
 compounds, allows very small moments to be measured with great
 accuracy, which is crucial for low fields and high
 temperatures. It 
 yields
microscopic information on the magnitude and orientation of the
field induced ordered moments.
 Our data analysis is based on the local susceptibility
 approach\cite{Gukasov93}. We show that it is valid in fields up
 to 1~T, and temperatures 5$<$$T$$<$270~K.
 It takes advantages of the high symmetry of the pyrochlore
 lattice which allows us to refine these complex magnetic structures with only two parameters,
 for any orientation of the magnetic field.

 For all temperatures and fields $\bf{H}$ // [110], we observe an F-like structure with propagation vector $\bf{k}$ = 0,
  characterized by magnetic Bragg peaks of the face centered cubic lattice.  We
 show that
 it involves $\alpha$ and
 $\beta$-chains 
  as in model spin
 ices. But in Tb$_2$Ti$_2$O$_7$, 
 these chains have very different moment values.
 This original effect is described quantitatively using the
 CF parameters derived in Ref.\onlinecite{Mirebeau07}.
 It provides a nice example of different moments induced by the
 field on the same crystallographic sites of a homogeneous and highly symmetric
 lattice.
\begin{figure}
    \includegraphics* [width=\columnwidth] {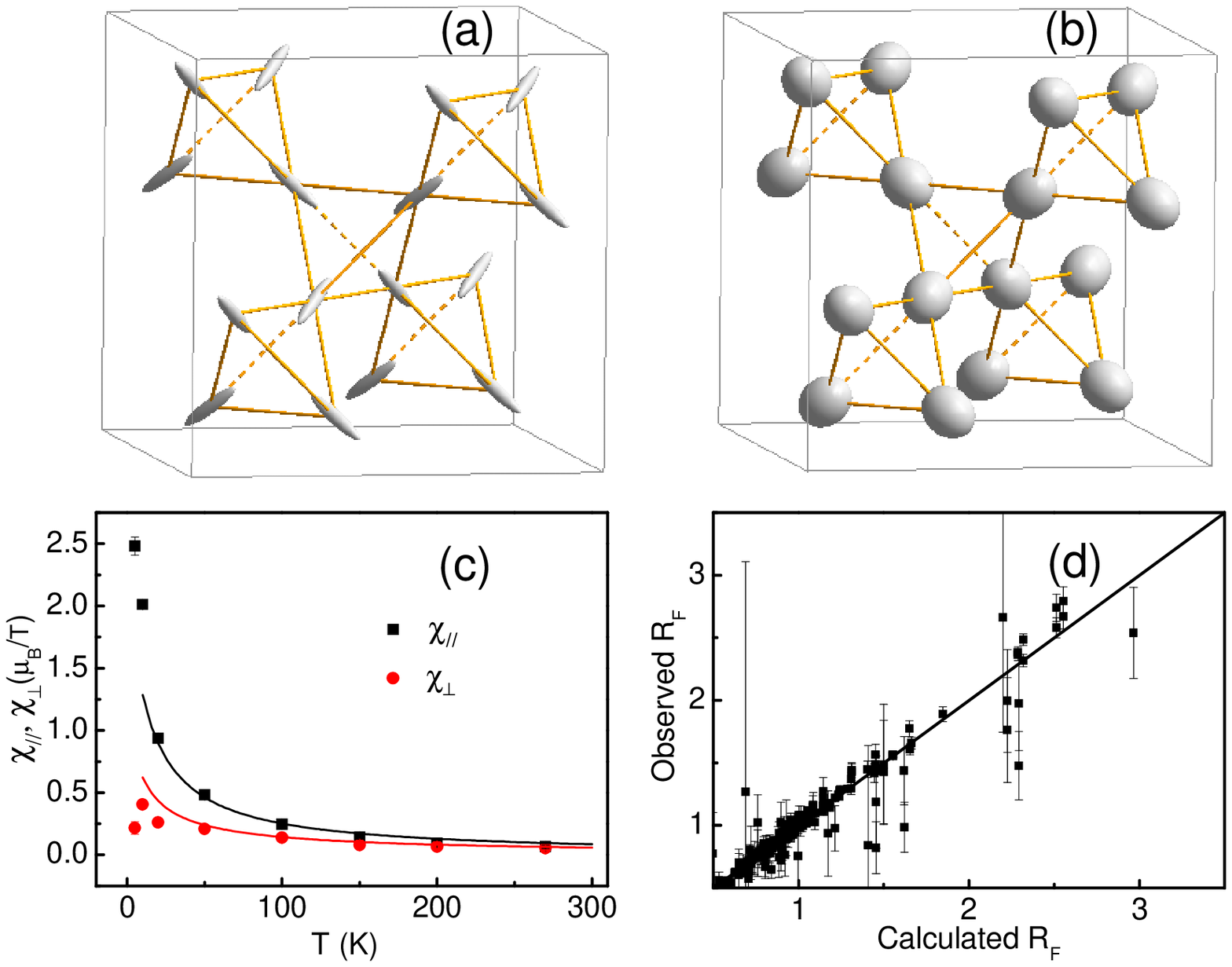}
    \caption {(Color online)\tbti: Local anisotropic susceptibility ellipsoids
    $\chi_{ij}$$T$, measured at 10~K (a) and 270~K (b). Ellipsoids
    were scaled by temperature to compensate the Curie behavior
    of the susceptibility.
   (c) susceptibility components $\chi_\varparallel$ and $\chi_\perp$ versus $T$. The lines are
calculations using the CF parameters of \tbti\ (see text); d: Measured versus calculated flipping ratio at 10~K; the refinement carried out with $\bf{H}$=1~T // [110] on 423 flipping ratios gave
$\chi_{11}$=0.939(15) $\mu_B$/T and $\chi_{12}$=0.535(11)
$\mu_B$/T with the goodness of fit $\chi^2$=2.92.
    }
    \label{Fig1}
\end{figure}

 Below 2~K and for fields above 2~T, we also observe a second family of Bragg
 peaks, belonging to the simple cubic lattice\cite{Rule06}.
 They are indexed in the cubic unit cell of $Fd\bar{3}m$ space group  with a propagation vector
 $\bf{k}$ = (0,0,1).
We studied this AF-like structure down to 0.3~K in fields up to 7
T, using unpolarized neutrons. We discuss its observation together
with the F-like structure and propose an explanation for its
origin. The whole analysis provides a microscopic description of
the field induced magnetic structures in Tb$_2$Ti$_2$O$_7$.

The neutron diffraction studies were performed on the
diffractometer Super-6T2  at the Orph\'ee reactor of the
Laboratoire L\'eon Brillouin \cite{Super6T2}. The field was
applied close to a $<$110$>$ axis, with a  misorientation of
$5^\circ$. We used unpolarized neutrons of wavelength
$\lambda_{n}=0.90$~\AA, collecting ~200 reflections for each ($H$,
$T$) set, for 10 temperatures between 0.3~K and 50~K and 7 fields
between 0 and 7~T. We also used polarized neutrons of incident
wavelength
$\lambda_{n}=0.84$~\AA ~ and polarization $P_0=0.91$. 
 We collected  flipping ratios at 200 to 400 Bragg peaks 
for each ($H$, $T$) set, for 8 temperatures
between 5 and 270~K, in a field of 1~T.
The programs FULLPROF \cite{Carvajal93}  and CHILSQ 
\cite{ccsl} were used to refine 
the magnetic intensities and flipping ratios, respectively.

A single crystal of \tbti was grown 
by the floating-zone technique. Its crystal structure was refined
at 50 K within the $Fd\bar{3}m$ 
 space group (434 reflections measured in zero-field), in agreement
 with powder data\cite{Han04}.

In the pyrochlore lattice, 
the 16 Tb atoms of the cubic unit cell occupy a single site (16$d$)
with local symmetry $\bar{3}m$. They can be subdivided into 4
different groups according to the directions of their local $<$111$>$ anisotropy axes. For $\bf{H}$
// [110] the $Fdd2$ group must be used in the magnetic moment
refinement\cite{Gukasov93}. It is the highest symmetry subgroup of
the $Fd\bar{3}m$ 
space group which leaves the magnetization 
invariant. In this subgroup the 16$d$ sites split into two different
sub-sets corresponding  to the $\alpha$-chains (where $\bf{H}$
makes an angle of $35.3^\circ$ with the local $<$111$>$ easy axis) and to
the $\beta$-chains (where it is perpendicular), respectively.

 In the local susceptibility approach\cite{Gukasov93},
 the atomic site susceptibility tensor accounts for the linear paramagnetic response of the
 moments to an applied field of arbitrary direction. The induced magnetic moment $\bf{M}^d$
 at the 16$d$ site of the unit cell writes: 
 $\bf{M^d}$ =
$\bar\bar\chi^d$$\bf{H}$, where $\bar\bar\chi^d$ is a tensor of rank 3x3,
whose components $\chi_{ij}$ depend on the symmetry of the atomic
site. The magnetic moments are in general not parallel to the
field, and the magnetic structure is not necessarily collinear, as
reflected
 by the non diagonal terms in the susceptibility
tensor. 
In the cubic axes, the symmetry constraints on the susceptibility tensor for a
magnetic atom at the 16$d$ site in the $Fd\bar 3m$ group imply:
$\chi_{11}=\chi_{22}=\chi_{33}$; $\chi_{12}=\chi_{13}=\chi_{23}$.
So only two independent parameters 
need to be determined regardless of the field direction. The
susceptibility components $\chi_\varparallel$ and $\chi_\perp$,
respectively parallel or perpendicular to the local $<$111$>$ easy
axis, measure the anisotropy on a given Tb ion. They are given by
the relations: $\chi_\varparallel$=$\chi_{11}$+2$\chi_{12}$ and
$\chi_\perp$=$\chi_{11}$-$\chi_{12}$.

We determined the susceptibility components by performing a least
square fit of the site susceptibility model to the flipping ratios
measured for $\bf{H}$=1~T // [110]. The refinement at 10~K (Fig. 1d) shows
the excellent quality of the fit.
 The magnetic ellipsoids determined from the site
susceptibility parameters are shown in Fig. 1a and 1b at 10~K and
270~K. For a given ellipsoid, the radius vector gives the magnitude and
direction of the magnetic moment induced by a field rotating in space.
The elongation of the ellipsoid increases with decreasing $T$,
reflecting the increasing CF anisotropy, as the magnetic moments
evolve from Heisenberg to Ising behavior, being more and more
constrained to align along their local $<$111$>$ axes. At 10~K the
two components $\chi_\varparallel$ and $\chi_\perp$ (Fig. 1c)
along the easy and hard local axes differ by a factor 5. They
still differ up to 200~K, but become practically equal at 270 K.
The lines in Fig.\ref{Fig1}c were calculated using the CF
interaction in \tbti\cite{Mirebeau07}, with a magnetic field of
1~T applied either parallel or perpendicular to the local $<111>$
axis. Good agreement with the data requires the AF exchange to be
taken into account: using a self-consistent calculation, valid
down to $\sim$10~K, we obtain a molecular field constant $\lambda
= -0.35$~T/\mub, close to the value derived in
Ref.\onlinecite{Mirebeau07}.


Once the site susceptibility parameters are known, one can easily
calculate the magnitude and the direction of the moments induced
on each Tb site by a field applied in an arbitrary direction. This
approach is valid for fields of 1~T  and temperatures
5$<$$T$$<$270~K, a range where the susceptibility parameters obey
a Curie-Weiss law (Fig. 1c).  Below $\sim$5~K, the local
susceptibility approach is no longer valid and we used unpolarized
neutron diffraction to determine the magnetic structure. We
refined the unpolarized neutron data assuming 4 different moments
(in magnitude and orientation) on the 4 independent Tb sites of
the cubic unit cell. The values plotted in Fig. 2 for $\bf{H}$ =
1~T
// [110] combine the analysis from polarized neutrons, for
5$<$$T$$<$270~K, and unpolarized neutrons for 0.3$<$$T$$<$10~K.
The two sets of data show a perfect overlap.

 For $\bf{H}$=1 T // [110], the $\bf{k}$ = 0 structure (Fig. 2a) at 0.3~K clearly shows a
  local spin ice order, with two moments pointing in and two out of a given
  tetrahedron. All moments keep close to their $<$111$>$ easy
  axis. The four tetrahedra of the unit cell are the same, as in Tb$_2$Sn$_2$O$_7$
  ordered spin ice\cite{Mirebeau05}. But in striking contrast with standard spin ice structures,
  here the Tb moments have very different magnitudes (Fig. 2b).
 The $\alpha$-moments reach 5.7(3) \mub at 0.3~K, close to the 5.7(2)
 \mub value deduced from the crystal field study\cite{Mirebeau07},
 whereas the $\beta$-moments only reach 1-2 $\mu_B$. The
 angular dependencies (not shown here), show that $\alpha$-moments
keep close to their local $<$111$>$ easy axis 
up to about 3~K, then
 reorient smoothly along $\bf{H}$ with increasing $T$. The $\beta$-moments 
 remain perpendicular to $\bf{H}$
  up to about 3~K, then reorient abruptly along $\bf{H}$.

  The presence of two types of $\beta$ moments ($\beta_1$ and
  $\beta_2$) with different values
  can be attributed to the misorientation of $\bf{H}$ with
respect to the [110] axis ($5^\circ$).  At 1~T and 0.3~K, our
refinements yield $\beta$-moment values of 0.8(2) and 1.3(2)
$\mu_B$. In contrast, the two types of
$\alpha$-moments 
keep almost the same values.

  The induced $\alpha$ and $\beta$-moments vary with temperature as
  predicted by the crystal field level scheme of \tbti \cite{Mirebeau07}, in the whole $T$
 range 0.3-270~K (Fig. 2b). In the calculation, the field of 1~T
 was taken exactly along [110] axis, so one needs to average the measured $\beta$-moments to compare
with the calculated ones. The agreement is quite good. It shows
that the non-zero $\beta$-moments found in the $\bf{k}$ = 0
structure, not seen in model spin ices\cite{Fennell05}, result
from the weaker Tb anisotropy and not from the field
misorientation. A more sophisticated model should involve both
applied and internal fields, and their precise orientations with
respect to the crystal axes. We note that different $\beta_1$ and
  $\beta_2$ moments are predicted by a CF calculation taking the field misorientation into account.
 At low
temperature, the strongly Ising character should render the
$\beta$-moments much more sensitive to the field orientation than
the $\alpha$-moments (Fig. 15 of ref. \onlinecite{Mirebeau07}), as
observed. Knowing the $\alpha$ and $\beta$-moments, one can also
 calculate the average magnetization $M$ in one tetrahedron, 
(Fig. 2b inset), which perfectly agrees with the bulk
magnetization data\cite{Mirebeau07}.



\begin{figure}
{\includegraphics* [width=\columnwidth] {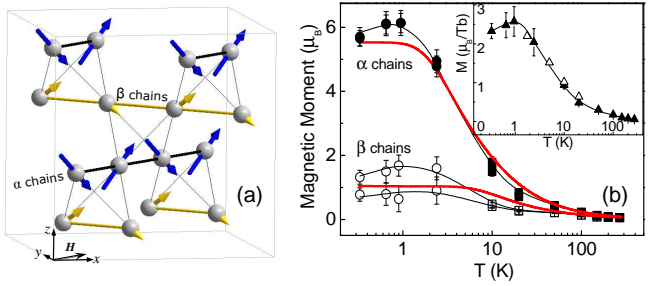}
 }
\caption{(Color online)\tbti: (a) field induced $\bf{k}$ = 0
 F-like structure at 0.3~K ($\bf{H}$=1T//[110]). Balls and arrows
represent Tb ions and moments, respectively. (b) temperature
dependence of the Tb moments from unpolarized ($\bullet$ for
$\alpha$-chains and $\circ$ for $\beta$-chains) and polarized
($\blacksquare$ for $\alpha$-chains and $\square$ for
$\beta$-chains) neutron data. The two $\beta$ moments arise from
the $5^\circ$ field misorientation. The thick red lines are a
crystal field calculation with a field $H$=1~T applied at
35.3$^\circ$ (resp. 90$^\circ$) from the local [111] axis for
$\alpha$ (resp. $\beta$) sites. Inset: magnetization $M$ versus
temperature: ($\blacktriangle$) from the neutron data,
($\triangle$) from bulk measurements of
Ref.\onlinecite{Mirebeau07}. Thin solid lines are guides to the
eye.
} \label{Fig2}
\end{figure}

\begin{figure}
\includegraphics* [width=\columnwidth] {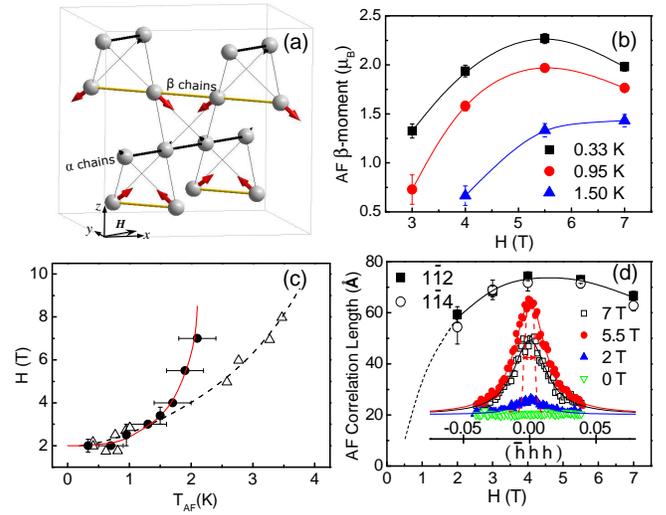}
\caption{(Color online)\tbti: (a) field induced $\bf{k}$ = (0,0,1)
AF-like structure
 at 0.3~K, ($\bf{H}$=7~T // [110]). (b) field dependence of the $\beta$
moments in the AF-like structure; the $\alpha$ moments (not
plotted) are below 0.2 $\mu_B$. (c) ($H,T_{AF}$) transition line
($\bullet$) our measurements, ($\triangle$)
 from Ref.\onlinecite {Rule06}. (d) AF correlation
length versus $H$; in inset, the (1$\bar{1}$2) peak for several
fields. Solid lines are fits with a Lorentzian peak shape, and
dashed line is the resolution peak shape.} \label{Fig3}
\end{figure}


\begin{figure}
\includegraphics* [width=\columnwidth] {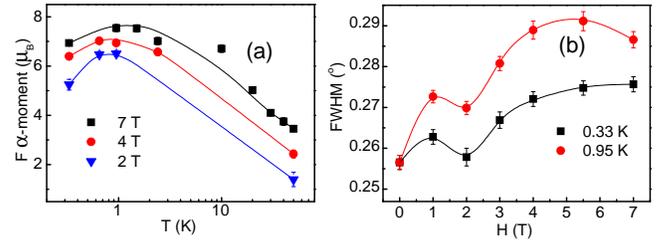}

\caption{(Color online)\tbti: (a) temperature dependence of the
$\alpha$-moments in the F-like structure in high fields ($H>$2~T).
The F $\alpha$-moments increase with $H$ and decrease with $T$
below $\sim$2~K when the AF-like structure settles in. (b) Field
dependence of the full width half maximum (FWHM) of the
(4$\bar{4}$0) peak with
negligible magnetic contribution. The value at $H$=0 is the resolution limit. 
} \label{Fig4}
\end{figure}

When $\bf{H}$ increases, the moment values in the $\bf{k}$ = 0
structure increase (Fig. 4a), and their angles with $\bf{H}$
decrease, showing that they reorient along the field. The F
$\alpha$-moments quickly saturate, whereas the F $\beta$-moments
increase more slowly with $\bf{H}$, reaching 3.2(3) \mub and
6.0(3) \mub for $\beta_1$ and $\beta_2$, respectively, at 7~T and
0.3~K.

  For $\bf{H}$ $>$ 2~T and below 2~K, new magnetic peaks appear,
  much weaker than the previous ones,
 indexed in the space
group $Fd\bar{3}m$ with
 $\bf{k}$=(0,0,1).
  We refined this new family of peaks as a second magnetic structure,
 also involving the whole sample, and analyzed independently. The $\bf{k}$=(0,0,1) value means that
 in the cubic cell, two tetrahedra have the same moment
 orientations and the other two are reversed, so that this AF-like
 structure has no net magnetization.
 In the $\bf{k}$=(0,0,1) structure, only the AF $\beta$-moments
 are significant, the AF $\alpha$-moments are
 negligible (Fig. 3a and 3b).
 At 0.3~K, up to $H$=7~T, the AF $\beta$-moments
 remain oriented at 10(5)$^\circ$ from their local $<$111$>$ axis,
 and 90(5)$^\circ$ from the field.

 The $\bf{k}$=(0,0,1) structure is stabilized in a certain ($H$, $T$) range, determined
 by plotting the peak integrated intensities versus $T$ or $H$.
 This yields 
 the transition line $T_{AF}$($H$) plotted in Fig. 3c. It
 agrees
 with a previous determination\cite{Rule06} for 0.3$<$$T$$<$1.5~K, but deviates above. 
The $\bf{k}$=(0,0,1) structure has a finite correlation length
$\xi_{AF}$, as shown by the Lorentzian
 linewidth of the magnetic peaks (Fig. 3d). $\xi_{AF}$ starts increasing
 with $H$, saturates at about 70 $\AA$  around 4~T, then
 decreases. Its variation reflects that of the AF $\beta$-moments (Fig. 3a), suggesting a reentrant
behavior. This AF-like structure should disappear at very high
fields, where all Tb moments are aligned.

We now summarize the field induced GS in Tb$_2$Ti$_2$O$_7$. Below
2~T, the GS consists of a LRO $\bf{k}$=0 structure, with field
induced moments both on $\alpha$ and $\beta$-chains. The moments
values are well explained by CF calculations. The Tb anisotropy,
much weaker than in model spin ices, yields non-zero moments on
the $\beta$-chains. Above 2~T, the GS is a superposition of two
modes with $\bf{k}$=0 and (0,0,1), the latter ordering with a
finite length scale. The $\bf{k}$= (0,0,1) structure occurs
together with a decrease of the F $\alpha$-moments in the
$\bf{k}$=0 structure (Fig. 4a) and the onset of well defined spin
waves\cite{Rule06}. It is also connected with a field broadening
of the $\it{nuclear}$ Bragg
  peaks (Fig. 4b), suggesting a lattice distortion. All these features
  reflect
  a symmetry breaking with respect to the quantum spin ice
  state\cite{Molavian07}
  stable at $H$=0.

   The $\bf{k}$= (0,0,1) structure induced by H//$\ $[110] strongly resembles that induced by a
  stress along the same direction\cite{Mirebeau04}. This strongly suggests
  that this structure is induced by magnetostriction effects. For ${H}$=0,
   a small distortion was observed in \tbti below 20~K, 
  likely precursor of a Jahn-Teller transition\cite{Ruff07}. In
  applied field,
   taking the bulk modulus\cite{Apetrei07} (B$_0$=156 GPa) and volume magnetostriction \cite{Alexandrov85}
    of Tb$_2$Ti$_2$O$_7$ at 4~K, we estimate that
 a field of 7~T is equivalent to
   a pressure of 0.03 GPa, well below the stress of 0.2 GPa that induces the AF order\cite{Mirebeau04}.
   However this estimation based on the isotropic volume is only a lower estimate for the stress. In
   Tb$_2$Ti$_2$O$_7$, magnetostriction coefficients $\varparallel$ and $\perp$ $\bf{H}$
   have very high values (up to 5. 10$^{-4}$), of opposite signs which
   nearly compensate in the isotropic volume magnetostriction.
   The strong anistropic magnetostriction arises from the specific Tb crystal
   field. The sample length $\varparallel$ $\bf{H}$ expands whereas
   that $\perp$ $\bf{H}$ contracts\cite{Alexandrov85}. At a microscopic scale,
   one could speculate that bonds in $\alpha$-chains
   expand, weakening the near neighbor exchange interaction $J$$_{nn}$ along these chains,
   whereas bonds in $\beta$-chains contract, reinforcing $J$$_{nn}$.
   This could explain the AF field induced periodicity along the
   $\beta$-chains.

    In conclusion,
    we determined the field induced magnetic structures in \tbti by combining polarized
    and unpolarized neutron diffraction, in a wide temperature and
     field range ($\bf{H}$ // [110]). The low field GS consists in a $\bf{k}$=0 (F-like) structure, with
     the local structure of a spin ice, but with Tb moments of very different
     magnitudes. With increasing field, the Tb moments reorient
     from the spin ice easy axes to the field axis.
    Above 2~T, a $\bf{k}$=(0,0,1) (AF-like)
    structure is also stabilized. It exists in a limited ($T$, $H$) range, has a finite length scale and shows reentrant behavior.
    It is attributed to magnetostriction. As temperature
increases, the susceptibility ellipsoids show a progressive change
from Ising to Heisenberg behavior. This change is demonstrated
using the local susceptibility approach, and well explained by the
 crystal field anisotropy of the Tb ion. 



\begin{thebibliography}{}
\bibitem{Bramwell012} S. T. Bramwell and M. J. P. Gingras, Science \textbf{294}, 1495 (2001).
\bibitem{Ramirez99} A. P. Ramirez {\em et al.} Nature \textbf{399}, 333 (1999). 
\bibitem{Harris98} M. J. Harris {\em et al.}  
Phys. Rev. Lett. \textbf{81}, 4496 (1998).
\bibitem{Matsuhira01} K. Matsuhira, Y.
Hinatsu, T. Sakakibara J. Phys.:Condens. Matter \textbf{13}, L737
(2001).
\bibitem{Sakakibara03} T. Sakakibara {\em et al.} 
Phys. Rev. Lett.
\textbf{90}, 207205 (2003).
\bibitem{Tabata06}Y. Tabata {\em et al.} 
Phys. Rev. Lett. \textbf{97},
257205, (2006). 
\bibitem{Castelnovo08} C. Castelnovo, R. Moessner, S. L. Sondhi, Nature \textbf{451},
42 (2008).
\bibitem{Jaubert08} L. D. C. Jaubert {\em et al.} 
Phys. Rev. Lett. \textbf{100}, 067207
(2008). 
\bibitem{Hiroi03} Z. Hiroi, K. Matsuhira and M. Ogata, J. Phys. Soc. Jpn. \textbf{72}, 3045 (2003).
\bibitem{Ruff05} J. P. C. Ruff, R. G. Melko, M. J. P. Gingras, Phys. Rev. Lett. \textbf{95}, 097202 (2005).
\bibitem{Fennell05} T. Fennell {\em et al.} 
Phys. Rev. B Phys. Rev. B \textbf{72},
224411 (2005).
\bibitem{Gardner99} J. S. Gardner {\em et al.}, 
 Phys. Rev. Lett. \textbf{82}, 1012 (1999).
\bibitem{Enjalran04} M. Enjalran and M. J. P. Gingras Phys. Rev. B \textbf{70}, 174426
(2004).
\bibitem{Molavian07} H. R. Molavian, M. J. P. Gingras, and B. Canals, Phys. Rev. Lett. \textbf{98}, 157204 (2007).
\bibitem{Mirebeau05} I. Mirebeau {\em et al.} 
Phys. Rev. Lett. \textbf{94}, 246402 (2005).
\bibitem{Mirebeau02} I. Mirebeau {\em et al.} 
Nature \textbf{420}, 54 (2002).
\bibitem{Mirebeau04} I. Mirebeau {\em et al.}
 Phys. Rev. Lett.
\textbf{93}, 187204 (2004).
\bibitem{Rule06} K. C. Rule {\em et al.}
 Phys. Rev. Lett. {\bf 96}, 177201 (2006). 
\bibitem{Gukasov93} A. Gukasov and P. J. Brown. J. Phys. Condens. Matt. {\bf 14}, 8831, (2002).
\bibitem{Mirebeau07} I. Mirebeau, P. Bonville, and M. Hennion, Phys. Rev. B. \textbf{76}, 184436 (2007). 
\bibitem{Super6T2} A. Gukasov {\em et al.} 
Physica B \textbf{397}, 131–134 (2007).
\bibitem{Carvajal93} J. Rodr\'{\i}guez-Carvajal, Physica B {\bf 192}, 55 (1993).
\bibitem{ccsl} P. J. Brown and J. C. Matthewman, CCSL-RAL-93-009 (1993), and http://www.ill.fr/dif/ccsl/html/ccsldoc.html
\bibitem{Ruff07} J. P. C. Ruff {\em et al.} 
Phys. Rev. Lett. \textbf{99}, 237202 (2007). 
\bibitem{Han04} S.-W. Han, J. S. Gardner, C. H. Booth, Phys. Rev. B {\bf 69}, 024416 (2004).
\bibitem{Alexandrov85} I. V. Aleksandrov {\em et al.} 
Sov. Phys. JETP \textbf{62}, 1287 (1985).
\bibitem{Apetrei07} A. Apetrei {\em et al.} 
J. Phys. Cond. Mat. \textbf{19}, 376208 (2007).





\end{thebibliography}
\end{document}